\documentclass[times, twoside]{zHenriquesLab-Style-arXiv}

\usepackage{balance}

\leadauthor{Zamora} 

\journal{Frontiers in Neuroscience}
\atdoi{10.3389/fnins.2021.734265}
\begin{document}

\title{Case Study: Embedding Digital Chronotherapy into Medical Devices – A canine validation for controlling status epilepticus through multi-scale rhythmic\\ brain stimulation}
\shorttitle{Embedding Digital Chronotherapy in Medical Devices}

% Use letters for affiliations, numbers to show equal authorship (if applicable) and to indicate the corresponding author
\author[1$\dag$\,\Letter]{Mayela Zamora}
\author[2$\dag$\,\Letter]{Sebastian Meller}
\author[3]{Filip Kajin}
\author[1,3]{James J. Sermon}
\author[3]{Robert Toth}
\author[1,3]{Moaad Benjaber}
\author[3]{Rafal Bogacz}
\author[4,5]{Derk-Jan Dijk}
\author[6]{Gregory A. Worrell}
\author[7]{Antonio Valent{\'i}n}
\author[3]{Benoit Duchet}
\author[2$\ddag$]{Holger A. Volk}
\author[1,3$\ddag$]{Timothy Denison}

\affil[1]{Institute of Biomedical Engineering, Department of Engineering Sciences, University of Oxford, Oxford OX3 7DQ, United Kingdom}
\affil[2]{Department of Small Animal Medicine and Surgery, University of Veterinary Medicine Hannover, 30559 Hannover, Germany}
\affil[3]{MRC Brain Network Dynamics Unit, Nuffield Department of Clinical Neurosciences, University of Oxford, Oxford OX1 3TH, United Kingdom}
\affil[4]{Surrey Sleep Research Centre, University of Surrey, Guildford GU2 7XP, United Kingdom}
\affil[5]{UK Dementia Research Institute, Care Research and Technology Centre, Imperial College London and the University of Surrey, Guildford GU2 7XP, United Kingdom}
\affil[6]{Department of Neurology, Mayo Clinic, Rochester, MN 55905, USA}
\affil[7]{Department of Clinical Neurophysiology, King's College Hospital NHS Trust, London SE5 9RS, United Kingdom}

\maketitle

%TC:break Abstract
%the command above serves to have a word count for the abstract
\begin{abstract}
Abstract --- Circadian and other physiological rhythms play a key role in both normal homeostasis and disease processes. Such is the case of circadian and infradian seizure patterns observed in epilepsy. However, these rhythms are not fully exploited in the design of active implantable medical devices. In this paper we explore a new implantable stimulator that implements chronotherapy as a feedforward input to supplement both open-loop and closed-loop methods. This integrated algorithm allows for stimulation to be adjusted to the ultradian, circadian and infradian patterns observed in patients through slowly-varying temporal adjustments of stimulation and algorithm sub-components, while also enabling adaption of stimulation based on immediate physiological needs such as a breakthrough seizure or change of posture. Embedded physiological sensors in the stimulator can be used to refine the baseline stimulation circadian pattern as a ``digital zeitgeber,'' i.e. a source of stimulus that entrains or synchronises the subject's natural rhythms. This algorithmic approach is tested on a canine with severe drug-resistant idiopathic generalized epilepsy exhibiting a characteristic diurnal pattern correlated with sleep-wake cycles. Prior to implantation, the canine's cluster seizures evolved to status epilepticus (SE) and required emergency pharmacological intervention. The cranially-mounted system was fully-implanted bilaterally into the centromedian nucleus of the thalamus. Using combinations of time-based modulation, thalamocortical rhythm-specific tuning of frequency parameters as well as fast-adaptive modes based on activity, the canine experienced no further SE events post-implant as of the time of writing (seven months). Importantly, no significant cluster seizures have been observed either, allowing the reduction of rescue medication. The use of digitally-enabled chronotherapy as a feedforward signal to augment adaptive neurostimulators could prove a useful algorithmic method in conditions where sensitivity to temporal patterns are characteristics of the disease state, providing a novel mechanism for tailoring a more patient-specific therapy approach.
\end{abstract}
%TC:break main
%the command above serves to have a word count for the abstract


\begin{keywords}
$^\dag$M.Z.~and~S.M.~contributed~equally~to~this~work~and~share~first~authorship.
\end{keywords}


\begin{corrauthor}
\newline {\normalfont \Letter} \space mayela.zamora@eng.ox.ac.uk
\newline {\normalfont \Letter} \space sebastian.meller@tiho-hannover.de
\end{corrauthor}

%%%%%%%%%%%%%%%%%%%%%
\section{Introduction}

Physiological rhythms play a role in both normal homeostasis and disease processes, yet the design of active implantable medical devices often does not fully exploit them, especially in brain stimulators. For example, in the treatment of epilepsy with deep brain stimulation (DBS), the default stimulation approach is to apply high-frequency (HF) stimulation in an attempt to suppress seizure propagation \cite{Fisher2010} -- a method adapted from the successful treatment of Parkinson's disease (PD). While beneficial in many cases, occasionally resulting in periods with seizure freedom \cite{Velasco2007, Valentin2013}, an exploration of alternative strategies, or a combination of strategies \cite{SchulzeBonhage2019} could give new insights for epilepsy treatment. Similar opportunities exist in other disease states such as movement disorders and neuropsychiatry. One approach is to exploit the precise digital time control of implantable systems to interact with the rhythmic processes in the brain.

Normal and pathological rhythms arise at multiple timescales. At one temporal extreme, $\SI{24}{\hour}$ (circadian) and multiday (infradian) seizure patterns are observed in epilepsy \cite{Leguia2021,Baud2018,Gregg2020}. Despite these temporal fluctuations, current FDA-approved DBS and responsive neurostimulation (RNS) devices for control of seizures run a fixed algorithm regardless of the time. Vagal nerve stimulators do enable two settings for implementing diurnal control, which show promise for managing side-effects and correlating therapy with symptoms \cite{Fisher2021}. Similarly, disrupted sleep-wake cycles are a common co-morbidity of PD, depression and epilepsy, but current DBS devices default to fixed, tonic stimulation parameters that are configured based on an assessment of efficacy during a daytime follow-up \cite{Malhotra2018}. At the faster end of the spectrum, thalamocortical oscillations are signatures of both healthy and diseased brain states that fluctuate in intensity on the order of tens to hundreds of milliseconds \cite{Oswal2013}. While these oscillations are used for adaptive algorithms, the stimulation paradigm is still largely reliant on gating HF stimulation for suppressing these lower frequency oscillations \cite{Little2013, Priori2013, Swann2018}. Stimulation at lower frequencies, utilizing oscillation frequencies recorded during natural behavior, however may provide additional benefits over HF stimulation due to the entrainment properties of the target neural population. DBS parameters could in principle be tuned to act as a ``digital chronotherapy'' that modulates endogenous rhythmicity in brain activity over multiple timescales.

In this case study, we apply a new implantable stimulator in the centromedian nucleus of the thalamus (CMN) that implements multi-scale, rhythm-entrained stimulation as an experimental medicine treatment for SE. For human generalized seizures, the CMN is involved early or late in the seizure and when involved, appears to lead the cortex \cite{MartinLopez2017}. Probably for this reason, this nucleus appears to be particularly useful for the treatment of super refractory SE in human patients \cite{Stavropoulos2021, Sa2019, Valentin2012}. SE is a serious ictal condition that is considered an emergent situation and can be fatal if these self-sustaining seizures cannot be interrupted. The stimulator's control algorithm applies feedforward input to supplement both open-loop and adaptive methods (Figure 1A). This integrated algorithm allows for electrical stimulation paradigms to be adjusted in response to slowly varying (e.g. diurnal/circadian) patterns through temporally-based adjustments of stimulation and algorithm sub-components, while also enabling adaptive stimulation based on immediate physiological needs such as a breakthrough seizure in epilepsy. The use of embedded field-potential sensing enabled subject-specific characterization of thalamocortical network activity. The field-potentials guided the application of targeted stimulation entrainment as an attempt to reinforce ``beneficial'' rhythms and avoid pathological ones. In aggregate, the physiological sensors and embedded timing control can serve to optimize the baseline stimulation circadian pattern as a digital zeitgeber, complementing or reinforcing existing zeitgebers.

%%%%%%%%%%%%%%%%%%%%%
\section{Case Overview}

A four-year-old, mixed-breed (Newfoundland / Saint Bernard), neutered male dog weighing $\SI{60}{\kilo\gram}$ was presented with severe drug-resistant idiopathic epilepsy, at the Tier II confidence level of diagnostic certainty \cite{DeRisio2015}. The carer's seizure diaries were used for comparative analysis of seizure type prevalence and frequency, seizure-free episodes and semiology of cluster seizures before and after surgery. The dog, treated as a veterinary patient, did not adequately respond to an array of antiseizure medication; treatment consisted of phenobarbital, potassium bromide, imepitoin, topiramate and gabapentin in various combinations (see Figure 4 for details). Multiple dosages of diazepam or levetiracetam as pulse therapy were used following any given seizure to prevent cluster seizures \cite{Packer2015}. The dog's diet was enriched with $\SI{6}{\percent}$ medium-chain triglycerides (MCT) with the goal to improve seizure control \cite{Berk2020}.

None of the epilepsy management options provided an adequate response and seizure severity increased to frequent SE. As no further medical treatment was available under the German Medicinal Products Act, the carer elected and gave informed consent for attempting DBS for epilepsy management. The Picostim--DyNeuMo research system (Bioinduction, Bristol, UK) was implanted with bilateral electrodes targeting the CMN, with the implantable pulse generator placed subcutaneously on the frontal cranium (Figure 1); refer to the supplemental methods for details. The Picostim--DyNeuMo can record intracranial signals and be remotely accessed for monitoring and therapy refinement; embedded circadian schedulers and sensors allow for adaptation of stimulation based on temporal patterns and inertial signals as well \cite{Toth2020}.

%%%%%%%%%%%%%%%%%%%%%
\section{Methods}

Initially after implantation, HF stimulation was used for stimulation consistent with prior reports of CMN stimulation ($\SI{130}{\hertz}$\,/\,$\SI{90}{\micro\second}$). However, in the first post-implant cluster seizures, increasing HF amplitudes led to intolerable side-effects without seizure cessation (head-pulling and other involuntary motion) and the cluster sequence proceeded unabated. This motivated the use of an analytical approach for low frequency entrainment.

\subsection{Theoretical Mechanism for Parameter Selection: Arnold Tongue Analysis for Estimation of Entrainment}

Our aim was to select stimulation frequencies which would reinforce neurotypical physiological behavior and avoid pathological rhythms. Prominent mesoscopic neural rhythms can be entrained through periodic electrical stimulation with specific amplitude and frequency predicted by Arnold tongues analysis. Entrainment may be subharmonic, characterized by a winding number $p:q$, with $p$ and $q$ integers, where $p$ is the average number of oscillations achieved by the rhythm for a given $q$ periodic pulses of the driving stimulation. Arnold tongues \cite{Arnold1983, Pikovsky2002} can be observed in the stimulation frequency/amplitude space as patterns of constant winding number, typically elongated and triangular in shape. The $p:q$ Arnold tongue represents the range of stimulation frequencies and amplitudes compatible with $p:q$ entrainment. Arnold tongues have previously been reported in computational models of brain circuits, in particular in the context of circadian rhythms \cite{Skeldon2017, Bordyugov2015} and transcranial stimulation \cite{Trevisan2006, Ali2013, Herrmann2016}.

The concept of Arnold tongues can be illustrated using the simplest model describing the influence of periodic stimulation on an oscillator. This model is the sine circle map \cite{Glass1979, Perez1982, Glass2001}, where a phase oscillator with constant natural frequency is forced by periodic stimulation of controlled frequency and amplitude. A stimulation pulse will advance or delay a neuron's phase depending on where the neuron is in its firing cycle and on the neuron's type \cite{Stiefel2008}. Similarly, stimulation in the sine circle map advances or delays the phase of the oscillator, such that the change in its phase is proportional to the sine of the oscillator's phase at the time of stimulation. Varying stimulation frequency and amplitude reveals a family of Arnold tongues as shown in Figure 2 for a natural frequency of $\SI{13}{\hertz}$. Highlighted in Figure 2A are the $1:1$ and $2:1$ tongues, which encompass stimulation parameters resulting in the oscillator frequency being entrained at exactly the stimulation frequency and at twice the stimulation frequency, respectively. Since the $1:1$ tongue is the largest, $1:1$ entrainment is the easiest to achieve in practice. For a fixed stimulation frequency, a broader range of natural frequencies can follow $1:1$ entrainment, which will therefore be most robust to perturbations acting to change the rhythm's natural frequency.

\subsection{Therapeutic Strategy -- Basal Stimulation Frequency and Fast Adaptation}

With remote telemetry, we were able to assess the spectral content of thalamocortical signals from our dataset based on prior characterization studies; representative power spectral density (PSD) plots are included in Figure 2B. Applying the entrainment hypothesis, we remotely tuned the stimulation frequency to the canine's dominant rhythm during restful, alert activity ($\SI{13}{\hertz}$\,/\,$\SI{350}{\micro\second}$\,/\,$\SI{1.3}{\milli\ampere}$ bilateral), while trying to avoid a sub-harmonic rhythm which might align with the $\SI{2}{\hertz}$ oscillation that correlated with seizure onset and initiation. Similar low frequency rhythms have also been suggested to induce absence seizures in human subjects \cite{Velasco1997}. The final entrainment model that guided therapy is summarized in the Arnold Tongue plot of Figure 2C. The adoption of this setting coincided with the end of the immediate cluster seizure event and it has been used thereafter as the default stimulation pattern. As an emergency fall-back for breakthrough seizures, a HF mode with elevated amplitude ($\SI{130}{\hertz}$\,/\,$\SI{90}{\micro\second}$\,/\,$\SI{1.5}{\milli\ampere}$ bilateral) was implemented which could be triggered by the carer through tap activation, using the built-in accelerometer (Figure 1B). Note that the levels for the emergency HF stimulation would not be tolerated during normal activities of daily living, e.g. it can induce reversible head-pulling, but were acceptable for an emergent state.

\subsection{Therapeutic Strategy -- Diurnal Rhythms and Slow Adaptation}

Since physiological rhythms can vary throughout the day \cite{Leguia2021, Gregg2020}, the stimulation might benefit from temporal adjustments regardless of immediate physiological state. In case of our canine stimulation could lead to hypervigilance, so the therapy was adjusted to vary over time, supplemented by adaptive transitions based on activity/inactivity. 

The temporal pattern to stimulation adjustment was introduced so as to align maximum stimulation intensity with times of peak seizure activity, as recorded in the seizure diary kept by the carer. The historic seizure activity up to the date of implantation is presented in a rose plot, inset in the right panel of Figure 3. The timing of seizures motivated a circadian adaptive pattern for stimulation; note that seizures were generally linked to sleep states according to the carer.

The aim was to account for immediate variation in activity, while accounting for daytime naps, since the highest probability of seizures correlated with the sleep state. The final adaptive algorithm, merging chronotherapy and sensor-based inputs, consisted of three layers of control with increasing stimulation intensity: 1) a circadian basal rate while the dog is awake and active; 2) a protective sleep mode with elevated entrainment stimulation; and 3) a high-amplitude, HF stimulation pattern to try and abort a breakthrough seizure through existing DBS methods. The embedded algorithm is illustrated by the circles enclosing the rose diagram in Figure 3. The inner ring of stimulation is the default state at $\SI{13}{\hertz}$ when the dog is active; the {\it night-time} activity uses elevated stimulation $\SI{0.7}{\milli\ampere}$ for additional protection, while the {\it daytime} stimulation is lowered to $\SI{0.5}{\milli\ampere}$ to avoid any side-effects of stimulation and conserve energy during low seizure probability intervals. When the accelerometer detects an inactive state for 4 minutes, the algorithm transitions to the middle ring of stimulation for the {\it sleeping mode}, which elevates stimulation amplitude to $\SI{1.3}{\milli\ampere}$ at $\SI{13}{\hertz}$ to provide greater entrainment during the increased risk of seizures during sleep. Finally, the outer ring, or {\it boost mode}, is designed for breakthrough seizures, activated by the carer with a single tap on the device programmed with a detection threshold of $\SI{7}{g}$ in the z-axis (orthogonal from the device plane). In this mode, a burst of $\SI{130}{\hertz}$, $\SI{1.5}{\milli\ampere}$ bipolar stimulation is provided to interrupt a sustained seizure.

We remotely synchronized the device for the longer infradian rhythms \cite{Baud2018}. Remote telemetric access allowed us to characterize physiology and reprogram the system in the home environment, as well as check battery levels and tissue-electrode interface impedances. On the left side of Figure 3, extended time (e.g. 2-week infradian) updates are provided over an encrypted internet link with a local password-protected application running on a surface tablet. The patient controller is used to wirelessly update the embedded stimulation parameters.

%%%%%%%%%%%%%%%%%%%%%
\section{Results}

\subsection{Prevention of Status Epilepticus}

The data is summarized in Figure 4 based on seizure diary and care plan summary. Figure 4A shows the seizure number and medication dosage per month since epilepsy onset, while Figure 4B shows the same data on a daily basis from seven months before until seven months after implantation and stimulation onset.

Status epilepticus: Three months before implantation, the seizure severity increased dramatically. Seizures regularly escalated into SE, with three occurring prior to surgery, requiring the use of rescue intervention. After implantation and the use of described stimulation patterns, no SE occurred.

Rescue medications: After implantation of the device, levetiracetam administration as pulse therapy, with major side-effects, was initially continued after seizure occurrence in order to further interrupt cluster seizure evolution or SE (repetitive administration every $\SI{8}{\hour}$ with successive dose reduction over several days). It was possible to successfully break the cluster seizure emergence or evolution in seven seizure occurrence periods (total of nine seizures) via stimulation only without administering levetiracetam as additional rescue medication in these periods (Figure 4B). Phenobarbital as chronic treatment was continued over the whole observation time after implantation with a dose reduction from $\SI{13.3}{}$ to $\SI{12.5}{\milli\gram\per\kilo\gram\per\day}$ in November 2020. The carer also stopped MCT supplementation and the other antiseizure medications.

Breakthrough seizure intervention: In terms of proactively interrupting ongoing seizures by the carer, the ``boost'' (HF burst) emergency mode disrupted $14$ seizures, while eight seizures continued after interruption attempt. Another eight seizures were not interrupted because they were noticed too late or the {\it boost mode} was deactivated at those time points. The success rate of the active interruption attempts was thus approximately $\SI{64}{\percent}$.

\subsection{Significant Trends for Reduction of Cluster Seizures}

General trends: The mean number of seizures during a seizure occurrence period, i.e. periods of isolated seizures (IS) or coherent cluster seizures (CS), as well as the mean duration of these periods ($\mathrm{IS} = \SI{0}{\hour}$; $\mathrm{CS}>\SI{0}{\hour}$) as a measurement for severity were assessed before and after start of low frequency (LF) entrainment (including preoperative seizures). Since all SE were part of a CS event, they were included for these measurements. The overall number of seizures within a seizure occurrence period since epilepsy onset was $4.67 \pm 5.99$ [mean $\pm$ SD, range $1-26$] with seizures occurring over a time period of $\SI{16.21}{} \pm \SI{21.35}{\hour}$ [mean $\pm$ SD, range $0-74.5$] per seizure occurrence period. Before the start of LF entrainment (including preoperative seizures), the number of seizures during an ictal period was $5.84 \pm 6.73$ [mean $\pm$ SD, range $1-26$] vs. $1.77 \pm 1.24$ [mean $\pm$ SD, range $1-5$] after LF entrainment started [$\mathrm{p} < 0.05$]. The time between the first and last seizure during a seizure occurrence period was $\SI{20.57}{} \pm \SI{23.42}{\hour}$ [mean $\pm$ SD, range $0-74.5$] vs. $\SI{5.48}{} \pm \SI{8.95}{\hour}$ [mean $\pm$ SD, range $0-24$] before and after start of LF entrainment (including preoperative seizures), respectively [$\mathrm{p} < 0.05$].

The graph in Figure 4A shows that the seizure number decreased in general without showing increased episode frequency, while the graph in Figure 4B shows that seizure episodes got more frequent, but less severe than before stimulation.

%%%%%%%%%%%%%%%%%%%%%
\section{Discussion}

Physiology generally merges feedforward (e.g. circadian) and feedback (e.g. homeostatic) control mechanisms. Implantable bioelectronic systems, while capable of precision timing and adaptive control, have not yet fully adopted a similar integrated control scheme. One reason is the complexity of additional control variables that might burden the clinician while configuring the system; ultimately an additional benefit must be demonstrated to justify the added complexity. However, many systems might yield immediate benefit by simply synchronizing stimulation modification to other diurnal variables such as medication timing. For example, fixed tonic stimulation of neural targets that couple into the reticular activating network have shown impact on sleep architecture \cite{Voges2015}. The fact that many areas of neuromodulation -- epilepsy, PD, chronic pain and depression -- have sleep co-morbidities also motivates an exploration of aligning stimulation with diurnal cycles to both enhance therapy and avoid side-effects \cite{Sladky2021}.

Alignment of rhythms at multiple scales requires a consideration of entrainment properties. We used the model of Arnold tongues from dynamic systems theory for selecting objectively the stimulation frequency. Arnold tongues can be useful for considering how stimulation might lead to non-linear effects which might not be intuitively predicted and have surprising side-effects. For example, PD patients can have half-harmonic locking of gamma rhythms (e.g. $\SI{65}{\hertz}$ peak) in response to $\SI{130}{\hertz}$ stimulation frequency \cite{Swann2016}. This non-linear mapping of brain stimulation to network oscillations might result inadvertently in reinforcing undesirable side-effects such as dyskinesia \cite{Swann2016}. Critically, these observations support the hypothesis that the conditions for Arnold tongues and subharmonic entrainment of the cortex are present with DBS of the basal ganglia. There is evidence that the alpha rhythm might also play a role in epilepsy \cite{Abela2019}; our strategy of attempting to entrain at a slightly higher frequency might also provide potential benefits, which warrants further investigation. For longer temporal scales such as circadian rhythms, the impact of stimulation as a ``digital zeitgeber'' might also result in additional phase shifts between existing zeitgebers (e.g. daylight or eating) and the endogenous circadian rhythm. Such phase shifts could either help restore sleep patterns, or create undesirable side effects, depending on the entrainment characteristics. Validating and applying these non-linear models of entrainment with additional clinical research might help to optimize the timing of stimulation at multiple temporal scales of physiology.

Finally, risk mitigations for novel adaptive stimulation methods must be considered. In the Picostim--DyNeuMo system, these mitigations include constraining the stimulation space to a predefined set of parameters screened by clinicians. In addition, we define a fallback program that a patient or carer can revert to in the case of issues arising with the adaptive mode. This action resets the system to open-loop stimulation, which is the default for most existing neuromodulation approach. An overview of the risk strategy method can be found in \cite{Gunduz2019}.

%%%%%%%%%%%%%%%%%%%%%
\section{Case Limitations}

This case report has several limitations. The study is of a single canine, which limits the statistical conclusions, but primarily serves as pilot validation of the implant technology. Although our results are consistent with recent human case studies \cite{Stavropoulos2021, Sa2019, Valentin2012} where the benefit of CMN thalamic stimulation at LF relative to HF was observed providing further support for the clinical value of thalamic LF stimulation, additional tests are needed. We adapted the stimulation based on physiological measurements and chose a higher frequency stimulation for entrainment. In addition, the application of experimental medicine prevented us from applying a self-control such as terminating treatment and assessing the impact on seizures. During the course of stimulation exploration, however, we were able to confirm that stimulation at $\SI{2}{\hertz}$ in the CMN increased the probability of seizures (induction $<\SI{24}{\hour}$ after setting) consistent with previous observations \cite{Velasco1997}. The case is ongoing and the reported results are limited to the first seven months of follow-up. Prior studies of epilepsy have shown changing efficacy over many years, although arguably for the better on average \cite{Nair2020}. In addition, we are relying on manual seizure diary which can be unreliable \cite{Ukai2021}; the strongest evidence we have are the SE events, which are severe enough to not be missed by the carer. Finally, the Picostim--DyNeuMo is limited by law to investigational device applications at this time.

%%%%%%%%%%%%%%%%%%%%%
\section{Summary}

The synchronization of brain stimulation to endogenous rhythms is an emerging concept for therapy optimization. The use of digitally-enabled chronotherapy as a feedforward signal to augment adaptive neurostimulators could prove to be a useful algorithmic method where sensitivity to temporal rhythms are characteristics of the disease state, tailoring a more patient-specific therapy approach. Computational models predicting Arnold tongues can also guide the design of patient-specific stimulation parameters, which has often been a heuristic process. In this proof-of-concept study, using a novel chronotherapy-enabled device in a canine with severe drug-resistant idiopathic epilepsy, these methods had favorable outcomes in terms of improving seizure semiology, reducing coherent cluster seizures and controlling (or avoiding) SE. The carer reports a reduced fear of seizures and improved personal quality of life based on the reduction of seizure severity with the stimulation. The adaptability of this approach allows for individualized therapies that are supported by emerging adaptive devices with both physiological sensing and chronotherapy capability. In addition, this report of DBS in canine epilepsy further highlights the possibility of using veterinary medicine as a vehicle to test new device and treatment paradigms \cite{Potschka2013}.

%%%%%%%%%%%%%%%%%%%%%
%% Mandatory sections

\section*{Author Contributions}
Conceptualization T.D. and H.A.V.;
Design of the study T.D., H.A.V. and D.J.D.;
Surgery and veterinary care S.M., H.A.V. and F.K.;
Algorithm definition T.D., D.J.D., G.W. and A.V.;
Algorithm implementation M.Z. and M.B.;
Data analysis S.M., M.Z., R.T., R.B., B.D. and J.J.S.;
Figures T.D., S.M., F.K., R.T., M.Z., B.D. and J.J.S.;
Writing original draft T.D., M.Z., S.M. and H.A.V.;
Writing, reviewing and editing -- all authors.

\section*{Funding}
M.Z. and J.S. were supported by the Medical Research Council grant MC\_UU\_00003/3, R.B. and B.D. by grants MC\_UU\_12024/5 and MC\_UU\_00003/1. T.D. was funded by the Royal Academy of Engineering.

\section*{Acknowledgments}
The authors would like to thank the directors and staff of Bioinduction for their invaluable help in the development and support of the Picostim--DyNeuMo research system. We also thank Theresia Henne and Anke Nagel for their support in the presurgical planning. Special thanks go to the patient's family for the dedication and care. We also extend our appreciation to Karen Wendt for her help translating during the follow up remote sessions with the carer.

\section*{Declaration of Interests}
The University of Oxford has research agreements with Bioinduction Ltd. T.D. also has business relationships with Bioinduction for research tool design and deployment. G.A.W. has financial interest in Cadence Neuroscience Inc. S.M., F.K., D.J.D., A.V. and H.V. declare no competing interests.
\vspace{1.0\baselineskip}
\balance

\section*{References}
\bibliography{main}

\begin{thebibliography}{41}
\providecommand{\natexlab}[1]{#1}
\providecommand{\url}[1]{\texttt{#1}}
\expandafter\ifx\csname urlstyle\endcsname\relax
  \providecommand{\doi}[1]{doi: #1}\else
  \providecommand{\doi}{doi: \begingroup \urlstyle{rm}\Url}\fi

\bibitem[Fisher et~al.(2010)Fisher, Salanova, Witt, Worth, Henry, Gross,
  Oommen, Osorio, Nazzaro, Labar, Kaplitt, Sperling, Sandok, Neal, Handforth,
  Stern, DeSalles, Chung, Shetter, Bergen, Bakay, Henderson, French, Baltuch,
  Rosenfeld, Youkilis, Marks, Garcia, Barbaro, Fountain, Bazil, Goodman,
  McKhann, Babu~Krishnamurthy, Papavassiliou, Epstein, Pollard, Tonder, Grebin,
  Coffey, Graves, and Group]{Fisher2010}
R.~Fisher, V.~Salanova, T.~Witt, R.~Worth, T.~Henry, R.~Gross, K.~Oommen,
  I.~Osorio, J.~Nazzaro, D.~Labar, M.~Kaplitt, M.~Sperling, E.~Sandok, J.~Neal,
  A.~Handforth, J.~Stern, A.~DeSalles, S.~Chung, A.~Shetter, D.~Bergen,
  R.~Bakay, J.~Henderson, J.~French, G.~Baltuch, W.~Rosenfeld, A.~Youkilis,
  W.~Marks, P.~Garcia, N.~Barbaro, N.~Fountain, C.~Bazil, R.~Goodman,
  G.~McKhann, K.~Babu~Krishnamurthy, S.~Papavassiliou, C.~Epstein, J.~Pollard,
  L.~Tonder, J.~Grebin, R.~Coffey, N.~Graves, and S.~S. Group.
\newblock Electrical stimulation of the anterior nucleus of thalamus for
  treatment of refractory epilepsy.
\newblock \emph{Epilepsia}, 51\penalty0 (5):\penalty0 899--908, 2010.
\newblock \doi{10.1111/j.1528-1167.2010.02536.x}.

\bibitem[Velasco et~al.(2007)Velasco, Velasco, Velasco, Jim{\'e}nez,
  Carrillo-Ruiz, and Castro]{Velasco2007}
F.~Velasco, A.~L. Velasco, M.~Velasco, F.~Jim{\'e}nez, J.~D. Carrillo-Ruiz, and
  G.~Castro.
\newblock Deep brain stimulation for treatment of the epilepsies: the
  centromedian thalamic target.
\newblock In \emph{Operative Neuromodulation}, pages 337--342. Springer Vienna,
  2007.
\newblock \doi{10.1007/978-3-211-33081-4_38}.

\bibitem[Valent{\'i}n et~al.(2013)Valent{\'i}n, Navarrete, E., R., C., M., L.,
  N., J., R., G., and Alarcon]{Valentin2013}
A.~Valent{\'i}n, Garc{\'i}a Navarrete, Chelvarajah E., Torres R., Navas C.,
  Vico M., Torres L., Pastor N., Selway J., Sola R., R.~G., and G.~Alarcon.
\newblock Deep brain stimulation of the centromedian thalamic nucleus for the
  treatment of generalized and frontal epilepsies.
\newblock \emph{Epilepsia}, 54\penalty0 (10):\penalty0 1823--1833, 2013.
\newblock \doi{10.1111/epi.12352}.

\bibitem[Schulze-Bonhage(2019)]{SchulzeBonhage2019}
A.~Schulze-Bonhage.
\newblock Long-term outcome in neurostimulation of epilepsy.
\newblock \emph{Epilepsy Behav}, 91:\penalty0 25--29, 2019.
\newblock \doi{10.1016/j.yebeh.2018.06.011}.

\bibitem[Leguia et~al.(2021)Leguia, Andrzejak, Rummel, Fan, Mirro, Tcheng, Rao,
  and Baud]{Leguia2021}
M.~G. Leguia, R.~G. Andrzejak, C.~Rummel, J.~M. Fan, E.~A. Mirro, T.~K. Tcheng,
  V.~R. Rao, and M.~O. Baud.
\newblock Seizure cycles in focal epilepsy.
\newblock \emph{JAMA Neurology}, 78\penalty0 (4):\penalty0 454--463, 2021.
\newblock \doi{10.1001/jamaneurol.2020.5370}.

\bibitem[Baud et~al.(2018)Baud, Kleen, Mirro, Andrechak, King-Stephens, Chang,
  and Rao]{Baud2018}
M.~O. Baud, J.~K. Kleen, E.~A. Mirro, J.~C. Andrechak, D.~King-Stephens, E.~F.
  Chang, and V.~R. Rao.
\newblock Multi-day rhythms modulate seizure risk in epilepsy.
\newblock \emph{Nature Communications}, 9\penalty0 (1):\penalty0 88, 2018.
\newblock \doi{10.1038/s41467-017-02577-y}.

\bibitem[Gregg et~al.(2020)Gregg, Nasseri, Kremen, Patterson, Sturges, Denison,
  Brinkmann, and Worrell]{Gregg2020}
N.~M. Gregg, M.~Nasseri, V.~Kremen, E.~E. Patterson, B.~K. Sturges, T.~J.
  Denison, B.~H. Brinkmann, and G.~A. Worrell.
\newblock Circadian and multiday seizure periodicities, and seizure clusters in
  canine epilepsy.
\newblock \emph{Brain Communications}, 2\penalty0 (1):\penalty0 fcaa008, 2020.
\newblock \doi{10.1093/braincomms/fcaa008}.

\bibitem[Fisher et~al.(2021)Fisher, DesMarteau, Koontz, Wilks, and
  Melamed]{Fisher2021}
B.~Fisher, J.~A. DesMarteau, E.~H. Koontz, S.~J. Wilks, and S.~E. Melamed.
\newblock Responsive vagus nerve stimulation for drug resistant epilepsy: A
  review of new features and practical guidance for advanced practice
  providers.
\newblock \emph{Frontiers in Neurology}, 11\penalty0 (1863), 2021.
\newblock \doi{10.3389/fneur.2020.610379}.

\bibitem[Malhotra(2018)]{Malhotra2018}
R.~K. Malhotra.
\newblock Neurodegenerative disorders and sleep.
\newblock \emph{Sleep Med Clin}, 13\penalty0 (1):\penalty0 63--70, 2018.
\newblock \doi{10.1016/j.jsmc.2017.09.006}.

\bibitem[Oswal et~al.(2013)Oswal, Brown, and Litvak]{Oswal2013}
A.~Oswal, P.~Brown, and V.~Litvak.
\newblock Synchronized neural oscillations and the pathophysiology of
  {Parkinson's} disease.
\newblock \emph{Curr Opin Neurol}, 26\penalty0 (6):\penalty0 662--670, 2013.
\newblock \doi{10.1097/WCO.0000000000000034}.

\bibitem[Little et~al.(2013)Little, Pogosyan, Neal, Zavala, Zrinzo, Hariz,
  Foltynie, Limousin, Ashkan, FitzGerald, Green, Aziz, and Brown]{Little2013}
S.~Little, A.~Pogosyan, S.~Neal, B.~Zavala, L.~Zrinzo, M.~Hariz, T.~Foltynie,
  P.~Limousin, K.~Ashkan, J.~FitzGerald, A.~L. Green, T.~Z. Aziz, and P.~Brown.
\newblock Adaptive deep brain stimulation in advanced {Parkinson} disease.
\newblock \emph{Annals of Neurology}, 74\penalty0 (3):\penalty0 449--457, 2013.
\newblock \doi{10.1002/ana.23951}.

\bibitem[Priori et~al.(2013)Priori, Foffani, Rossi, and Marceglia]{Priori2013}
A.~Priori, G.~Foffani, L.~Rossi, and S.~Marceglia.
\newblock Adaptive deep brain stimulation {(aDBS)} controlled by local field
  potential oscillations.
\newblock \emph{Exp Neurol}, 245:\penalty0 77--86, 2013.
\newblock \doi{10.1016/j.expneurol.2012.09.013}.

\bibitem[Swann et~al.(2018)Swann, de~Hemptinne, Thompson, Miocinovic, Miller,
  Gilron, Ostrem, Chizeck, and Starr]{Swann2018}
N.~C. Swann, C.~de~Hemptinne, M.~C. Thompson, S.~Miocinovic, A.~M. Miller,
  R.~Gilron, J.~L. Ostrem, H.~J. Chizeck, and P.~A. Starr.
\newblock Adaptive deep brain stimulation for {Parkinson's} disease using motor
  cortex sensing.
\newblock \emph{J Neural Eng}, 15\penalty0 (4):\penalty0 046006, 2018.
\newblock \doi{10.1088/1741-2552/aabc9b}.

\bibitem[Mart{\'i}n-L{\'o}pez et~al.(2017)Mart{\'i}n-L{\'o}pez,
  Jim{\'e}nez-Jim{\'e}nez, Caba{\~n}{\'e}s-Mart{\'i}nez, Selway, and
  Valent{\'i}n]{MartinLopez2017}
D.~Mart{\'i}n-L{\'o}pez, D.~Jim{\'e}nez-Jim{\'e}nez,
  L.~Caba{\~n}{\'e}s-Mart{\'i}nez, R.~P. Selway, and A.~Valent{\'i}n.
\newblock The role of thalamus versus cortex in epilepsy: Evidence from human
  ictal centromedian recordings in patients assessed for deep brain
  stimulation.
\newblock \emph{International Journal of Neural Systems}, 27\penalty0
  (07):\penalty0 1750010, 2017.
\newblock \doi{10.1142/S0129065717500101}.

\bibitem[Stavropoulos et~al.(2021)Stavropoulos, Selway, Hasegawa, Hughes,
  Rittey, J{\'i}menez-J{\'i}menez, and Valent{\'i}n]{Stavropoulos2021}
I.~Stavropoulos, R.~Selway, H.~Hasegawa, E.~Hughes, C.~Rittey,
  D.~J{\'i}menez-J{\'i}menez, and A.~Valent{\'i}n.
\newblock Low frequency centromedian thalamic nuclei deep brain stimulation for
  the treatment of super refractory status epilepticus: A case report and a
  review of the literature.
\newblock \emph{Brain Stimul}, 14\penalty0 (2):\penalty0 226--229, 2021.
\newblock \doi{10.1016/j.brs.2020.12.013}.

\bibitem[Sa et~al.(2019)Sa, Singh, Pujar, D'Arco, Desai, Eltze, Hughes, Obaidi,
  M., D., M., R., H., Kaliakatsos, and Valent{\'i}n]{Sa2019}
M.~Sa, R.~Singh, S.~Pujar, F.~D'Arco, N.~Desai, C.~Eltze, E.~Hughes, Al~Obaidi,
  Eleftheriou M., Tisdall D., Selway M., Cross R., J.~H., M.~Kaliakatsos, and
  A.~Valent{\'i}n.
\newblock Centromedian thalamic nuclei deep brain stimulation and anakinra
  treatment for {FIRES} - two different outcomes.
\newblock \emph{Eur J Paediatr Neurol}, 23\penalty0 (5):\penalty0 749--754,
  2019.
\newblock \doi{10.1016/j.ejpn.2019.08.001}.

\bibitem[Valent{\'i}n et~al.(2012)Valent{\'i}n, Nguyen, Skupenova,
  Agirre-Arrizubieta, Jewell, Mullatti, Moran, Richardson, Selway, and
  Alarc{\'o}n]{Valentin2012}
A.~Valent{\'i}n, H.~Q. Nguyen, A.~M. Skupenova, Z.~Agirre-Arrizubieta,
  S.~Jewell, N.~Mullatti, N.~F. Moran, M.~P. Richardson, R.~P. Selway, and
  G.~Alarc{\'o}n.
\newblock Centromedian thalamic nuclei deep brain stimulation in refractory
  status epilepticus.
\newblock \emph{Brain Stimul}, 5\penalty0 (4):\penalty0 594--598, 2012.
\newblock \doi{10.1016/j.brs.2011.10.002}.

\bibitem[De~Risio et~al.(2015)De~Risio, Bhatti, Mu{\~n}ana, Penderis, Stein,
  Tipold, Berendt, Farqhuar, Fischer, Long, Mandigers, Matiasek, Packer,
  Pakozdy, and Patterson]{DeRisio2015}
L.~De~Risio, S.~Bhatti, K.~Mu{\~n}ana, J.~Penderis, V.~Stein, A.~Tipold,
  M.~Berendt, R.~Farqhuar, A.~Fischer, S.~Long, P.~J.~J. Mandigers,
  K.~Matiasek, R.~M.~A. Packer, A.~Pakozdy, and N.~and Patterson.
\newblock International veterinary epilepsy task force consensus proposal:
  diagnostic approach to epilepsy in dogs.
\newblock \emph{BMC Veterinary Research}, 11\penalty0 (1):\penalty0 148, 2015.
\newblock \doi{10.1186/s12917-015-0462-1}.

\bibitem[Packer et~al.(2015)Packer, Nye, Porter, and Volk]{Packer2015}
R.~M.~A. Packer, G.~Nye, S.~E. Porter, and H.~A. Volk.
\newblock Assessment into the usage of levetiracetam in a canine epilepsy
  clinic.
\newblock \emph{BMC Veterinary Research}, 11\penalty0 (1):\penalty0 25, 2015.
\newblock \doi{10.1186/s12917-015-0340-x}.

\bibitem[Berk et~al.(2020)Berk, Law, Packer, Wessmann, Bathen-N{\"o}then,
  Jokinen, Knebel, Tipold, Pelligand, Meads, and Volk]{Berk2020}
B.~A. Berk, T.~H. Law, R.~M.~A. Packer, A.~Wessmann, A.~Bathen-N{\"o}then,
  T.~S. Jokinen, A.~Knebel, A.~Tipold, L.~Pelligand, Z.~Meads, and H.~A. Volk.
\newblock A multicenter randomized controlled trial of medium-chain
  triglyceride dietary supplementation on epilepsy in dogs.
\newblock \emph{Journal of Veterinary Internal Medicine}, 34\penalty0
  (3):\penalty0 1248--1259, 2020.
\newblock \doi{10.1111/jvim.15756}.

\bibitem[Toth et~al.(2020)Toth, Zamora, Ottaway, Gillbe, Martin, Benjaber,
  Lamb, Noone, Taylor, Deli, Kremen, Worrell, Constandinou, de~Wachter,
  Knowles, Sharott, Valentin, Green, and Denison]{Toth2020}
R.~Toth, M.~Zamora, J.~Ottaway, T.~Gillbe, S.~Martin, M.~Benjaber, G.~Lamb,
  T.~Noone, B.~Taylor, A.~Deli, V.~Kremen, G.~Worrell, T.~G. Constandinou,
  S.~de~Wachter, C.~Knowles, A.~Sharott, A.~Valentin, A.~L. Green, and
  T.~Denison.
\newblock {DyNeuMo Mk-2:} an investigational circadian-locked neuromodulator
  with responsive stimulation for applied chronobiology.
\newblock \emph{In Proceedings of the 2020 IEEE International Conference on
  Systems, Man, and Cybernetics (SMC)}, pages 3433--3440, 2020.
\newblock \doi{10.1109/SMC42975.2020.9283187}.

\bibitem[Arnol'd(1983)]{Arnold1983}
V.~I. Arnol'd.
\newblock Remarks on the perturbation theory for problems of {Mathieu} type.
\newblock \emph{Russian Mathematical Surveys}, 38:\penalty0 215, 1983.
\newblock \doi{10.1070/RM1983v038n04ABEH004210}.

\bibitem[Pikovsky et~al.(2002)Pikovsky, Rosenblum, and Kurths]{Pikovsky2002}
A.~Pikovsky, M.~Rosenblum, and J.~Kurths.
\newblock Synchronization: A universal concept in nonlinear science.
\newblock \emph{Am J Phys}, 70\penalty0 (6):\penalty0 655--655, 2002.
\newblock \doi{10.1119/1.1475332}.

\bibitem[Skeldon et~al.(2017)Skeldon, Phillips, and Dijk]{Skeldon2017}
A.~C. Skeldon, A.~J. Phillips, and D.~J. Dijk.
\newblock The effects of self-selected light-dark cycles and social constraints
  on human sleep and circadian timing: a modeling approach.
\newblock \emph{Sci Rep}, 7:\penalty0 45158, 2017.
\newblock \doi{10.1038/srep45158}.

\bibitem[Bordyugov et~al.(2015)Bordyugov, Abraham, Granada, Rose, Imkeller,
  Kramer, and Herzel]{Bordyugov2015}
G.~Bordyugov, U.~Abraham, A.~Granada, P.~Rose, K.~Imkeller, A.~Kramer, and
  H.~Herzel.
\newblock Tuning the phase of circadian entrainment.
\newblock \emph{J R Soc Interface}, 12\penalty0 (108):\penalty0 20150282, 2015.
\newblock \doi{10.1098/rsif.2015.0282}.

\bibitem[Trevisan et~al.(2006)Trevisan, Mindlin, and Goller]{Trevisan2006}
M.~A. Trevisan, G.~B. Mindlin, and F.~Goller.
\newblock Nonlinear model predicts diverse respiratory patterns of birdsong.
\newblock \emph{Phys Rev Lett}, 96\penalty0 (5):\penalty0 058103, 2006.
\newblock \doi{10.1103/PhysRevLett.96.058103}.

\bibitem[Ali et~al.(2013)Ali, Sellers, and Frohlich]{Ali2013}
M.~M. Ali, K.~K. Sellers, and F.~Frohlich.
\newblock Transcranial alternating current stimulation modulates large-scale
  cortical network activity by network resonance.
\newblock \emph{J Neurosci}, 33\penalty0 (27):\penalty0 11262--11275, 2013.
\newblock \doi{10.1523/JNEUROSCI.5867-12.2013}.

\bibitem[Herrmann et~al.(2016)Herrmann, Murray, Ionta, Hutt, and
  Lefebvre]{Herrmann2016}
C.~S. Herrmann, M.~M. Murray, S.~Ionta, A.~Hutt, and J.~Lefebvre.
\newblock Shaping intrinsic neural oscillations with periodic stimulation.
\newblock \emph{Journal of Neuroscience}, 36\penalty0 (19):\penalty0
  5328--5337, 2016.
\newblock \doi{10.1523/JNEUROSCI.0236-16.2016}.

\bibitem[Glass and Mackey(1979)]{Glass1979}
L.~Glass and M.~C. Mackey.
\newblock A simple model for phase locking of biological oscillators.
\newblock \emph{Journal of Mathematical Biology}, 7\penalty0 (4):\penalty0
  339--352, 1979.
\newblock \doi{10.1007/BF00275153}.

\bibitem[Perez and Glass(1982)]{Perez1982}
R.~Perez and L.~Glass.
\newblock Bistability, period doubling bifurcations and chaos in a periodically
  forced oscillator.
\newblock \emph{Phys Lett A}, 90\penalty0 (9):\penalty0 441--443, 1982.
\newblock \doi{10.1016/0375-9601(82)90391-7}.

\bibitem[Glass(2001)]{Glass2001}
L.~Glass.
\newblock Synchronization and rhythmic processes in physiology.
\newblock \emph{Nature}, 410\penalty0 (6825):\penalty0 277--284, 2001.
\newblock \doi{10.1038/35065745}.

\bibitem[Stiefel et~al.(2008)Stiefel, Gutkin, and Sejnowski]{Stiefel2008}
K.~M. Stiefel, B.~S. Gutkin, and T.~J. Sejnowski.
\newblock Cholinergic neuromodulation changes phase response curve shape and
  type in cortical pyramidal neurons.
\newblock \emph{PLoS One}, 3\penalty0 (12):\penalty0 e3947, 2008.
\newblock \doi{10.1371/journal.pone.0003947}.

\bibitem[Velasco et~al.(1997)Velasco, Velasco, Velasco, Brito, Jimenez,
  Marquez, and Rojas]{Velasco1997}
M.~Velasco, F.~Velasco, A.~L. Velasco, F.~Brito, F.~Jimenez, I.~Marquez, and
  B.~Rojas.
\newblock Electrocortical and behavioral responses produced by acute electrical
  stimulation of the human centromedian thalamic nucleus.
\newblock \emph{Electroencephalogr Clin Neurophysiol}, 102\penalty0
  (6):\penalty0 461--471, 1997.
\newblock \doi{10.1016/s0013-4694(96)95203-0}.

\bibitem[Voges et~al.(2015)Voges, Schmitt, Hamel, House, Kluge, Moll, and
  Stodieck]{Voges2015}
B.~R. Voges, F.~C. Schmitt, W.~Hamel, P.~M. House, C.~Kluge, C.~K. Moll, and
  S.~R. Stodieck.
\newblock Deep brain stimulation of anterior nucleus thalami disrupts sleep in
  epilepsy patients.
\newblock \emph{Epilepsia}, 56\penalty0 (8):\penalty0 e99--e103, 2015.
\newblock \doi{10.1111/epi.13045}.

\bibitem[Sladky et~al.(2021)Sladky, Nejedly, Mivalt, Brinkmann, Kim, Louis, K.,
  Gregg, Lundstrom, Crowe, Attia, Crepeau, Balzekas, Marks, Wheeler, Cimbalnik,
  Cook, Janca, Sturges, Leyde, Miller, Van~Gompel, Denison, Worrell, and
  Kremen]{Sladky2021}
V.~Sladky, P.~Nejedly, F.~Mivalt, B.~H. Brinkmann, I.~Kim, St. Louis, E.~K.,
  N.~M. Gregg, B.~N. Lundstrom, C.~M. Crowe, T.~P. Attia, D.~Crepeau,
  I.~Balzekas, V.~Marks, L.~P. Wheeler, J.~Cimbalnik, M.~Cook, R.~Janca, B.~K.
  Sturges, K.~Leyde, K.~J. Miller, J.~J. Van~Gompel, T.~Denison, G.~A. Worrell,
  and V.~Kremen.
\newblock Distributed brain co-processor for neurophysiologic tracking and
  adaptive stimulation: Application to drug resistant epilepsy.
\newblock \emph{bioRxiv}, 2021.
\newblock \doi{10.1101/2021.03.08.434476}.

\bibitem[Swann et~al.(2016)Swann, de~Hemptinne, Miocinovic, Qasim, Wang, Ziman,
  Ostrem, Luciano, M., B., and Starr]{Swann2016}
N.~C. Swann, C.~de~Hemptinne, S.~Miocinovic, S.~Qasim, S.~S. Wang, N.~Ziman,
  J.~L. Ostrem, San Luciano, Galifianakis M., N.~B., and P.~A. Starr.
\newblock Gamma oscillations in the hyperkinetic state detected with chronic
  human brain recordings in {Parkinson's} disease.
\newblock \emph{J Neurosci}, 36\penalty0 (24):\penalty0 6445--6458, 2016.
\newblock \doi{10.1523/JNEUROSCI.1128-16.2016}.

\bibitem[Abela et~al.(2019)Abela, Pawley, Tangwiriyasakul, Yaakub, Chowdhury,
  Elwes, Brunnhuber, and Richardson]{Abela2019}
Eugenio Abela, Adam~D. Pawley, Chayanin Tangwiriyasakul, Siti~N. Yaakub,
  Fahmida~A. Chowdhury, Robert D.~C. Elwes, Franz Brunnhuber, and Mark~P.
  Richardson.
\newblock Slower alpha rhythm associates with poorer seizure control in
  epilepsy.
\newblock \emph{Ann Clin Transl Neurol}, 6\penalty0 (2):\penalty0 333--43,
  2019.
\newblock \doi{10.1002/acn3.710}.

\bibitem[Gunduz et~al.(2019)Gunduz, Opri, Gilron, Kremen, Worrell, Starr,
  Leyde, and Denison]{Gunduz2019}
A.~Gunduz, E.~Opri, R.~Gilron, V.~Kremen, G.~Worrell, P.~Starr, K.~Leyde, and
  T.~Denison.
\newblock Adding wisdom to 'smart' bioelectronic systems: a design framework
  for physiologic control including practical examples.
\newblock \emph{Bioelectronics in Medicine}, 2\penalty0 (1):\penalty0 29--41,
  2019.

\bibitem[Nair et~al.(2020)Nair, Laxer, Weber, Murro, Park, Barkley, Smith,
  Gwinn, Doherty, Noe, Zimmerman, Bergey, Anderson, Heck, Liu, Lee, Sadler,
  Duckrow, Hirsch, Wharen, Tatum, Srinivasan, McKhann, Agostini, Alexopoulos,
  Jobst, Roberts, Salanova, Witt, Cash, Cole, Worrell, Lundstrom, Edwards,
  Halford, Spencer, Ernst, Skidmore, Sperling, Miller, Geller, Berg, Fessler,
  Rutecki, Goldman, Mizrahi, Gross, Shields, Schwartz, Labar, Fountain, Elias,
  Olejniczak, Villemarette-Pittman, Eisenschenk, Roper, Boggs, Courtney, Sun,
  Seale, Miller, Skarpaas, and Morrell]{Nair2020}
D.~R. Nair, K.~D. Laxer, P.~B. Weber, A.~M. Murro, Y.~D. Park, G.~L. Barkley,
  B.~J. Smith, R.~P. Gwinn, M.~J. Doherty, K.~H. Noe, R.~S. Zimmerman, G.~K.
  Bergey, W.~S. Anderson, C.~Heck, C.~Y. Liu, R.~W. Lee, T.~Sadler, R.~B.
  Duckrow, L.~J. Hirsch, R.~E. Wharen, Jr., W.~Tatum, S.~Srinivasan, G.~M.
  McKhann, M.~A. Agostini, A.~V. Alexopoulos, B.~C. Jobst, D.~W. Roberts,
  V.~Salanova, T.~C. Witt, S.~S. Cash, A.~J. Cole, G.~A. Worrell, B.~N.
  Lundstrom, J.~C. Edwards, J.~J. Halford, D.~C. Spencer, L.~Ernst, C.~T.
  Skidmore, M.~R. Sperling, I.~Miller, E.~B. Geller, M.~J. Berg, A.~J. Fessler,
  P.~Rutecki, A.~M. Goldman, E.~M. Mizrahi, R.~E. Gross, D.~C. Shields, T.~H.
  Schwartz, D.~R. Labar, N.~B. Fountain, W.~J. Elias, P.~W. Olejniczak, N.~R.
  Villemarette-Pittman, S.~Eisenschenk, S.~N. Roper, J.~G. Boggs, T.~A.
  Courtney, F.~T. Sun, C.~G. Seale, K.~L. Miller, T.~L. Skarpaas, and M.~J.
  Morrell.
\newblock Nine-year prospective efficacy and safety of brain-responsive
  neurostimulation for focal epilepsy.
\newblock \emph{Neurology}, 95\penalty0 (9):\penalty0 e1244--e1256, 2020.
\newblock \doi{10.1212/WNL.0000000000010154}.

\bibitem[Ukai et~al.(2021)Ukai, Parmentier, Cortez, Fischer, Gaitero, Lohi,
  Nykamp, Jokinen, Powers, Sammut, Sanders, Tai, Wielaender, and
  James]{Ukai2021}
Masayasu Ukai, Thomas Parmentier, Miguel~A. Cortez, Andrea Fischer, Luis
  Gaitero, Hannes Lohi, Stephanie Nykamp, Tarja~S. Jokinen, Danielle Powers,
  Veronique Sammut, Sean Sanders, Tricia Tai, Franziska Wielaender, and Fiona
  James.
\newblock Seizure frequency discrepancy between subjective and objective ictal
  electroencephalography data in dogs.
\newblock \emph{J Vet Intern Med}, 2021.
\newblock \doi{10.1111/jvim.16158}.

\bibitem[Potschka et~al.(2013)Potschka, Fischer, von Ruden, Hulsmeyer, and
  Baumgartner]{Potschka2013}
H.~Potschka, A.~Fischer, E.~L. von Ruden, V.~Hulsmeyer, and W.~Baumgartner.
\newblock Canine epilepsy as a translational model?
\newblock \emph{Epilepsia}, 54\penalty0 (4):\penalty0 571--579, 2013.
\newblock \doi{10.1111/epi.12138}.

\end{thebibliography}


\begin{thebibliography}{4}
\providecommand{\natexlab}[1]{#1}
\providecommand{\url}[1]{\texttt{#1}}
\expandafter\ifx\csname urlstyle\endcsname\relax
  \providecommand{\doi}[1]{doi: #1}\else
  \providecommand{\doi}{doi: \begingroup \urlstyle{rm}\Url}\fi

\bibitem[Toth et~al.(2020)Toth, Zamora, Ottaway, Gillbe, Martin, Benjaber,
  Lamb, Noone, Taylor, Deli, Kremen, Worrell, Constandinou, de~Wachter,
  Knowles, Sharott, Valent{\'i}n, Green, and Denison]{Toth2020}
R.~Toth, M.~Zamora, J.~Ottaway, T.~Gillbe, S.~Martin, M.~Benjaber, G.~Lamb,
  T.~Noone, B.~Taylor, A.~Deli, V.~Kremen, G.~Worrell, T.~G. Constandinou,
  S.~de~Wachter, C.~Knowles, A.~Sharott, A.~Valent{\'i}n, A.~L. Green, and
  T.~Denison.
\newblock {DyNeuMo Mk-2:} an investigational circadian-locked neuromodulator
  with responsive stimulation for applied chronobiology.
\newblock \emph{In Proceedings of the 2020 IEEE International Conference on
  Systems, Man, and Cybernetics (SMC)}, pages 3433--3440, 2020.
\newblock \doi{10.1109/SMC42975.2020.9283187}.

\bibitem[Velasco et~al.(2002)Velasco, Velasco, Jim{\'e}nez, Velasco, Rojas, and
  Perez]{Velasco2002}
Francisco Velasco, Marcos Velasco, Fiacro Jim{\'e}nez, Ana~Luisa Velasco,
  Beatriz Rojas, and Martha~Luisa Perez.
\newblock Centromedian nucleus stimulation for epilepsy: Clinical,
  electroencephalographic, and behavioral observations.
\newblock \emph{Thalamus \& Related Systems}, 1\penalty0 (4):\penalty0
  387--398, 2002.
\newblock \doi{10.1016/S1472-9288(02)00011-0}.

\bibitem[Valent{\'i}n et~al.(2013)Valent{\'i}n, Navarrete, E., R., C., M., L.,
  N., J., R., G., and Alarcon]{Valentin2013}
A.~Valent{\'i}n, Garc{\'i}a Navarrete, Chelvarajah E., Torres R., Navas C.,
  Vico M., Torres L., Pastor N., Selway J., Sola R., R.~G., and G.~Alarcon.
\newblock Deep brain stimulation of the centromedian thalamic nucleus for the
  treatment of generalized and frontal epilepsies.
\newblock \emph{Epilepsia}, 54\penalty0 (10):\penalty0 1823--1833, 2013.
\newblock \doi{10.1111/epi.12352}.

\bibitem[Cukiert and Leht{\"i}maki(2017)]{Cukiert2017}
Arthur Cukiert and Kai Leht{\"i}maki.
\newblock Deep brain stimulation targeting in refractory epilepsy.
\newblock \emph{Epilepsia}, 58\penalty0 (S1):\penalty0 80--84, 2017.
\newblock \doi{10.1111/epi.13686}.

\end{thebibliography}

\clearpage

%%%%%%%%%%%%%%%%%%%%%
%% Figures

\begin{figure*}[!h]
\centering
\includegraphics[width=180mm]{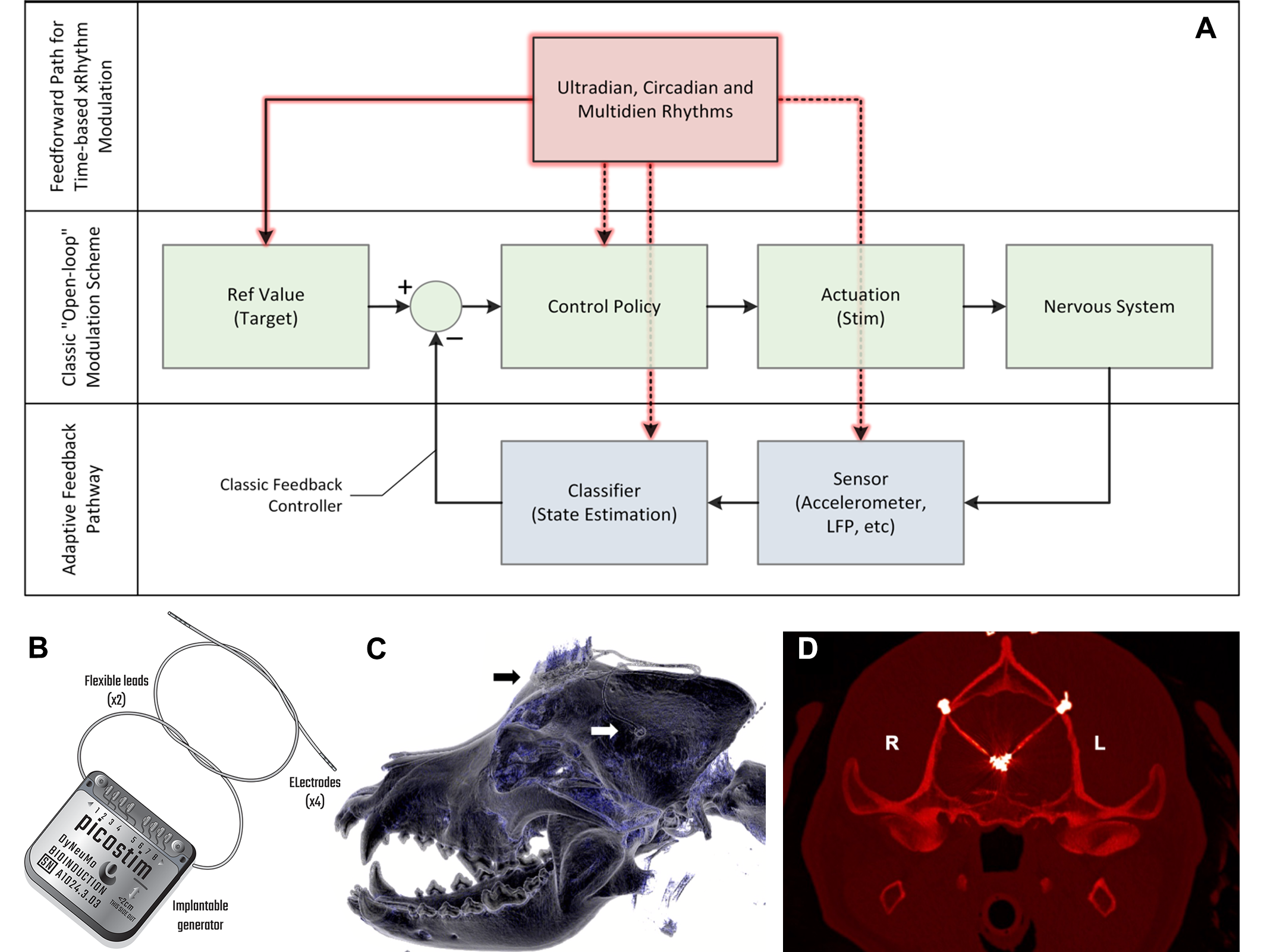}
\caption{(A) The model for the control algorithm in the device. The typical ``open loop'' pathway is illustrated in green. A clinician generally captures a reference value and control policy based on experience and observation, setting a stimulation state for the device that interacts with the nervous system. A ``closed loop'' pathway, illustrated in blue, can be constructed by adding a sensor and classifier that can then adjust stimulation accordingly through a control policy; this adaptive feedback pathway allows for continuous refinement based on immediate patient state. A feedforward pathway, illustrated in red, can help to account for variability linked to set temporal rhythms, similar to circadian cycles observed in physiology control pathways. The feedforward pathway can adjust multiple parameters in the open and adaptive loops, including reference values for measured variables, how to stimulate, which sensors to use and the classification objectives. The combination of feedforward and feedback pathways aims to optimize predictive and responsive stimulation control. (B) The Picostim--DyNeuMo cranial mounted deep brain stimulator used in this study. For device capabilities refer to \cite{Toth2020}. Note that the research tool is upgradeable through the firmware and software versions. (C) Postsurgical imaging shows a reconstructed 3D computed tomography (CT) scan of the canine's skull with the guide tube hub fixed in the drilled hole in the left parietal bone (white arrow) and the stimulator fixed on the frontal bone (black arrow). Wires coming from both guide tubes are connected with the stimulator. (D) Implanted guide tubes and wires in a transversal CT scan in the plane of the target structures. Note that the strong hyperintense signals around the target positions represent CT metal artefacts deriving from the electrode plates ($\mathrm{n} = 4$, each) at the tip region of the wires. Please see supplemental methods for more details.}
\end{figure*}

\begin{figure*}[!t]
\centering
\includegraphics[width=180mm]{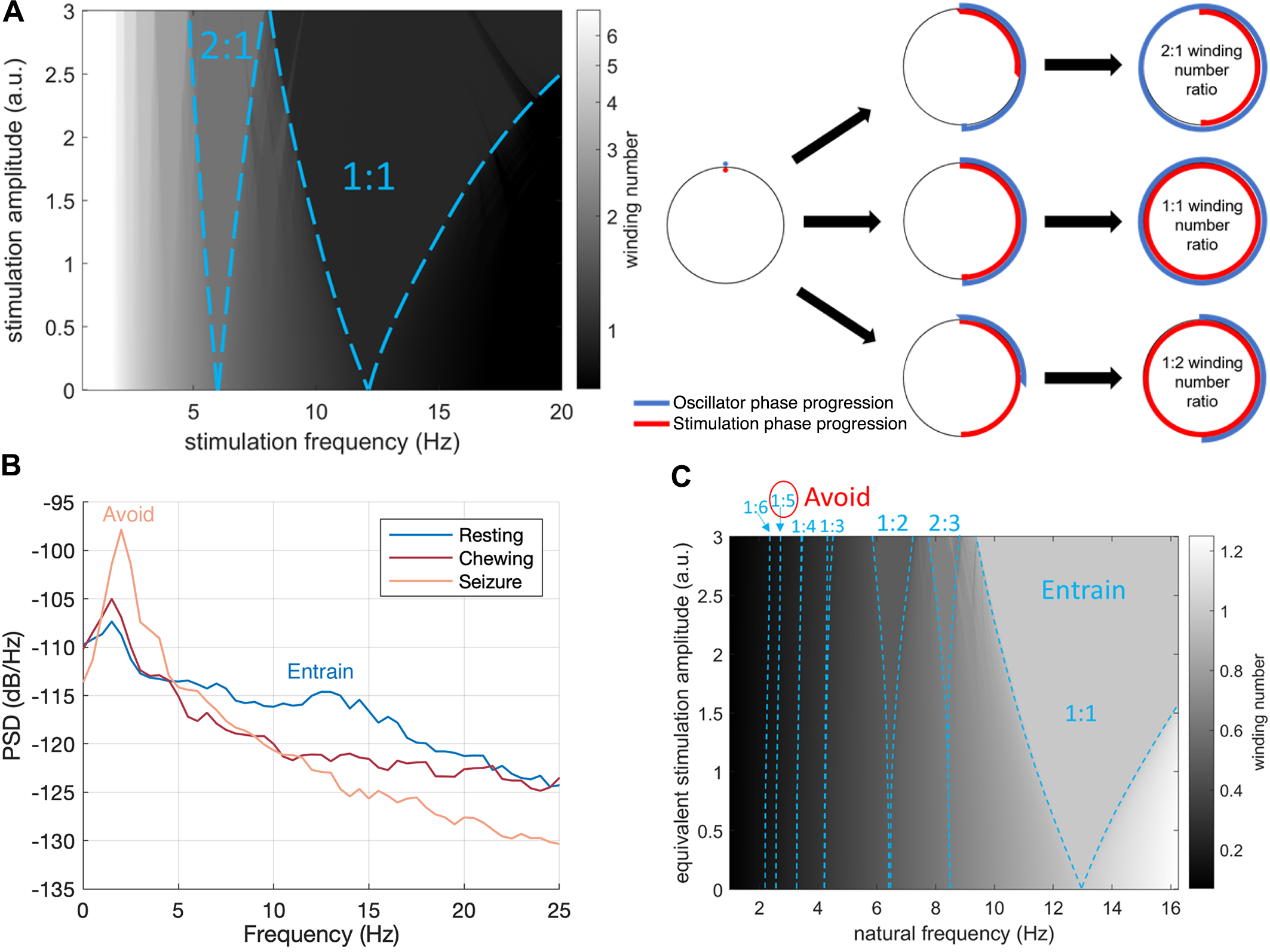}
\caption{(A) Winding number and Arnold tongues in the sine circle map as a function of stimulation (driver) frequency and amplitude for an oscillator with a natural frequency of $\SI{13}{\hertz}$. Arnold tongues correspond to areas of constant winding number. The $1:1$ tongue (winding number of 1) and the $2:1$ tongue (winding number of 2) are highlighted with blue dashed lines. For stimulation parameters falling within the $p:q$ tongue, the rhythm will be entrained at $p/q$ times the stimulation frequency. The sine circle map was simulated as $\theta_{i+1} = \theta_{i} + 2 \pi \left(f_{0} / f_{s}\right) + I sin\left(\theta_{i}\right)$, where $\theta_{i}$ is the oscillator phase right after stimulation pulse $i$, $f_{0}$ is the natural frequency of the oscillator, $f_{s}$ is the stimulation frequency and $I$ is the stimulation amplitude. The winding number was calculated after $N=50$ stimulation pulses as the average of $\left(\theta_{N} - \theta_{0}\right) / \left(2 \pi N\right)$ over $20$ trials with random initial phases $\theta_{0}$ uniformly distributed between $0$ and $2 \pi$. (B) Intracranial field potentials from the implanted signals (left hemisphere, contacts $0-3$) remotely accessed through wireless telemetry. Representative signals were gathered during different activities of daily living to characterize frequency content. The stimulation therapy strategy aims to entrain the healthy rhythm around $\SI{13}{\hertz}$, while avoiding the peak at $\SI{2}{\hertz}$ observed during seizure. (C) Illustration of the final stimulation strategy. The winding number in the sine circle map is shown here for a fixed stimulation frequency ($\SI{13}{\hertz}$) as a function of natural frequency (e.g. inherent thalamocortical rhythm) and stimulation amplitude. Selected Arnold tongues are highlighted in blue. Stimulation at $\SI{13}{\hertz}$ can reliably entrain the desired $\SI{12}{\hertz}$ thalamocortical oscillation (large $1:1$ tongue) while avoiding induction of pathological tongues in the region of $2-\mathrm{to}-\SI{3}{\hertz}$. The $1:5$ and $1:6$ tongues obtained from $\SI{13}{\hertz}$ stimulation are indeed so narrow that they will not lead to any entrainment in practice. To account for the fact that neural oscillations at lower frequencies typically have higher power ($1/f$ power law) and would therefore require more energy to entrain, the vertical axis represents equivalent stimulation amplitude (stimulation amplitude multiplied by $f_{0} / f_{max}$, where $f_{0}$ is the natural frequency and $f_{max}$ the maximum frequency shown). This is conservative as the $1:5$ and $1:6$ tongues disappear at higher stimulation amplitudes.}
\end{figure*}

\begin{figure*}[!h]
\centering
\includegraphics[width=180mm]{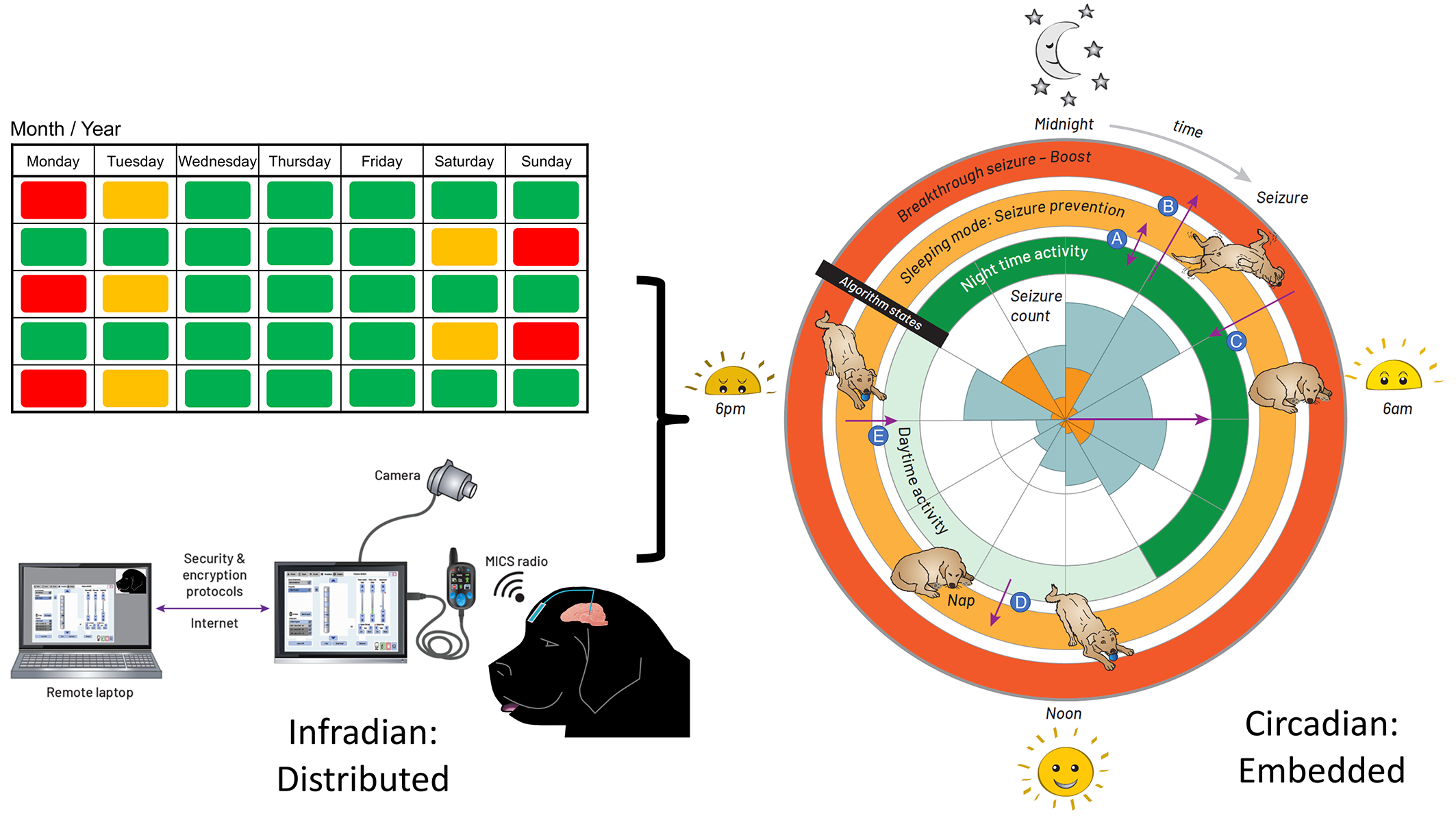}
\caption{Illustration of the algorithm timing strategy. Left: extended time (e.g. 2-week infradian) updates are provided over an encrypted internet link with a local password-protected application running on a surface tablet via the patient controller. The USB connector between the patient programmer and tablet is for in-clinic programming. Research subjects use the handheld controller for at-home recharge and manual adjustments. Right: the embedded algorithm illustrated with a rose plot. The inner circle represents the seizure count from the diary, kept by the carer, up to the date of DBS activation; the orange tiling is the timing of first seizure onset, while the blue account for all seizures in a cluster. The inner ring of stimulation is the {\it default state} when the dog is active. When the accelerometer detects an inactive state for 4 minutes, the algorithm transitions to the middle ring of stimulation for the {\it sleeping mode}. The outer ring is the {\it boost mode} for breakthrough seizures. (A) going in and out of sleep during the night, while in night-time mode, the stimulation switches from night-time to the sleeping mode and vice versa. (B) suffering a seizure during the default mode, the carer taps on the site of the implant to trigger the boost mode. (C) the stimulation returns from boost mode to the default mode within 30 minutes. (D) taking a nap during the day, the stimulation switches from the default mode to the sleeping mode. (E) waking up from the nap, the stimulation goes from sleeping mode to default mode within 30 minutes.}
\end{figure*}

\begin{figure*}[!t]
\centering
\includegraphics[width=180mm]{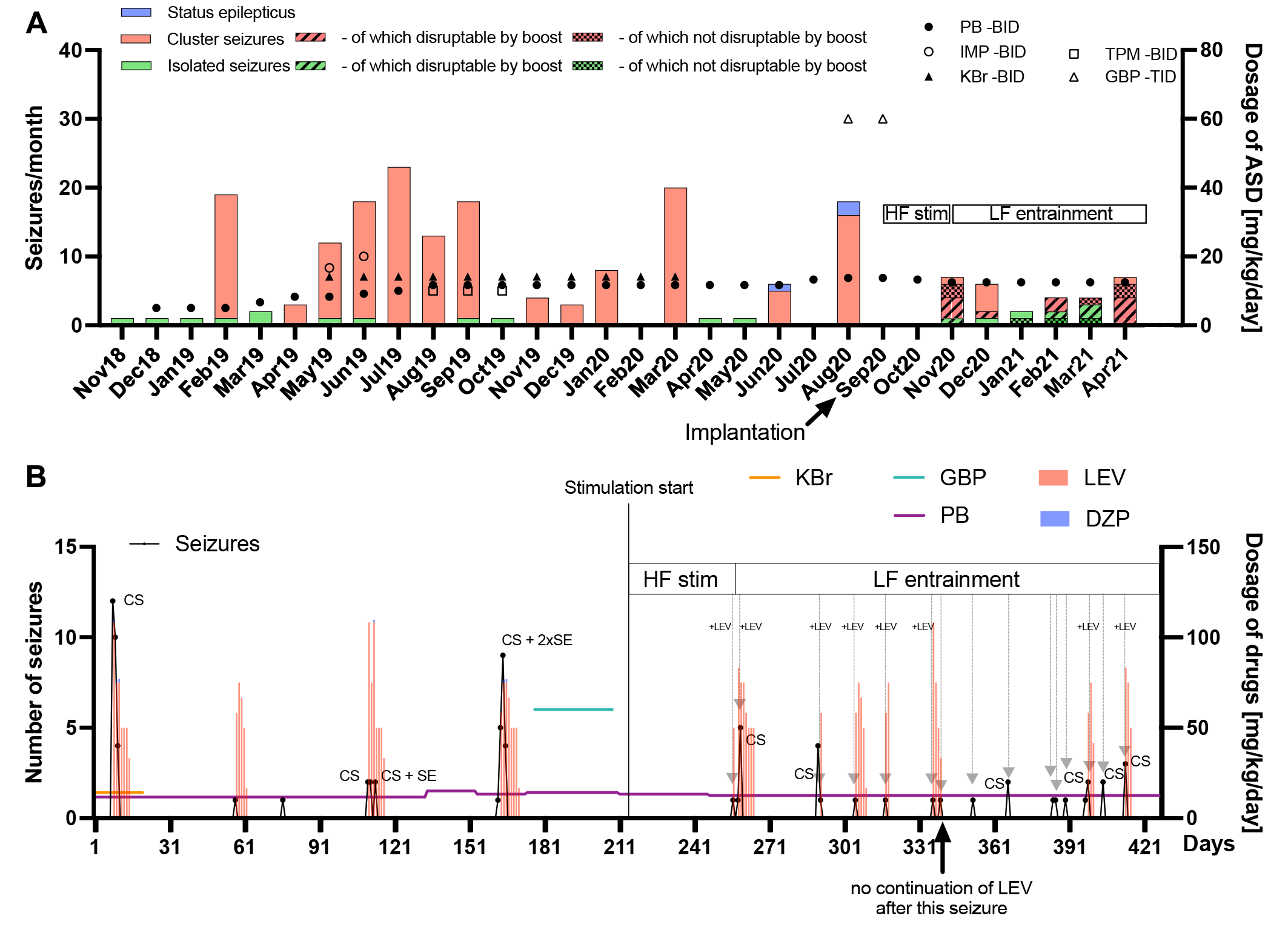}
\caption{(A) The seizure frequency (left y-axis) and average daily antiseizure drug (ASD) dosage in $\SI{}{\milli\gram\per\kilo\gram\per\day}$ (right y-axis) of phenobarbital (PB), imepitoin (IMP), potassium bromide (KBr), topiramate (TPM) and gabapentin (GBP) twice (BID) or thrice (TID) daily are shown for every month since epilepsy onset in November 2018. When stimulation was started, no further status epilepticus (SE) occurred. After implantation, high frequency (HF) stimulation was applied after seizures occurred in order to prevent SE or further cluster seizure evolution. Since HF did not bring the desired result in the first cluster seizure (November 2020), the frequency was subsequently changed to low frequency (LF) entrainment adapted to the canine's local field potentials. Phenobarbital (PB) was chronically administered as monotherapy (reduction from $\SI{13.3}{}$ to $\SI{12.5}{\milli\gram\per\kilo\gram\per\day}$ in November 2020), while other chronic medical and dietary therapy with medium-chain triglycerides was stopped after surgery. With the HF bursting `boost mode' for the interruption of ongoing seizures four seizures out of six attempts in November 2020 were interrupted. One seizure was noticed too late. In December 2020, one seizure was interrupted while five further ones occurred without interruption attempt. In January 2021, one attempt of seizure interruption was without success, while another seizure was noticed too late. In February 2021, two out of three attempts of seizure interruption were successful. In March 2021, three seizures were interrupted while two continued after attempt. In April 2021, four seizures were interrupted, two continued and one was noticed too late. The overall success rate of interruption attempts was $\SI{64}{\percent}$. Electrical stimulation has the potential of reducing acute and chronic pharmacological interventions. (B) The seizure number (left y-axis) in black dots and lines and average daily ASD dosage in $\SI{}{\milli\gram\per\kilo\gram\per\day}$ (right y-axis) for chronic (horizontal colored lines) and emergency treatment (vertical bars) with levetiracetam (LEV) or diazepam (DZP) are shown on a day-to-day basis since begin of March 2020 until end of April 2021 (seven months pre and post implantation). Grey arrows represent successful avoidance of further seizure occurrence/evolution after a single seizure or in cluster seizures (CS) via stimulation with (+LEV) or without LEV intervention. It turned out that initially, only the combination of acute LEV treatment and LF-entrainment prevented or interrupted cluster seizure evolution. Since February 2021, however, LEV was not administered after seizure occurrence due to severe side-effects, with LF-entrainment alone achieving cluster cessation, except in two CS events which consisted of rapid successive but few seizures in April 2021. Eight seizure evolution interruptions were conducted together with LEV and seven without LEV after stimulation onset, with only seven interruptions with LEV during LF-entrainment. DBS has the potential of reducing acute pharmacological interventions.}
\end{figure*}

\end{document}

\title{Case Study: Embedding Digital Chronotherapy into\\ Medical Devices – A canine validation for controlling status epilepticus through multi-scale rhythmic brain stimulation\\Supplementary Material}

\author{Mayela Zamora, Sebastian Meller, Filip Kajin, James J. Sermon, Robert Toth, Moaad Benjaber, Rafal Bogacz, Derk-Jan Dijk, Gregory A. Worrell, Antonio Valent{\'i}n, Benoit Duchet, Holger A. Volk, Timothy Denison}

\onecolumn

\maketitle

\begin{keywords}
% Left here for spacing
\end{keywords}

\begin{corrauthor}
\newline {\normalfont \Letter} \space mayela.zamora@eng.ox.ac.uk
\newline {\normalfont \Letter} \space sebastian.meller@tiho-hannover.de
\end{corrauthor}

%%%%%%%%%%%%%%%%%%%%%
\section{The Device}

The Picostim--DyNeuMo system (Bioinduction, Bristol, UK), serves as a $\SI{7}{\cm^3}$ cranially-mounted brain pacemaker, supplemented with on-demand telemetric monitoring of neurobiological signals and flexible and remotely adjustment of stimulation parameters to the patient's individual needs. Details of the device are provided in \cite{Toth2020}. Feedback responses enabled by biological inputs from the built-in accelerometers and from telemetric measurements of neuronal signals together with high resolution temporal adjustments to stimulation parameters represent key features. Furthermore, feedforward paths can be configured by e.g. using clock and activity measurement, or manipulations, via accelerometry. This provides a highly flexible system which offers an investigational hub for neuroscience and for the development of new open and closed loop deep brain stimulation (DBS) solutions.

\section{Presurgical Planning}

Both centromedian nuclei (CMN) of the thalamus were chosen as targets, as its electrical stimulation was shown to be effective in people with refractory generalized epilepsy syndromes \cite{Velasco2002, Valentin2013, Cukiert2017}. The StrykerNAV3\textsuperscript{\tiny\textregistered} hardware and Stryker CranialMap\textsuperscript{\tiny\textregistered} 3.0 software (Stryker Corporation, Kalamazoo, MI, USA) were used to perform presurgical planning and neuronavigation during surgery with related patient- and instrument-tracking devices. Magnetic resonance imaging and computed tomography (CT) images of the canine patient's skull and brain were merged for surgical planning of the bilateral tracks for the implantable Renishaw\textsuperscript{\tiny\textregistered} neuroguide  DBS electrode guide tubes (Renishaw plc., Gloucestershire, UK) and electrode placement. Bilateral trajectories ($\SI{36}{\mm}$ each from the skull surface) were planned through the ectosylvian gyrus of the temporal lobe from a rostro-dorso-lateral to caudo-ventro-medial direction with entry points in the rostral portion of the parietal bones and passing through and ending slightly ventromedially from the CMN of the corresponding hemisphere in order to ensure electrode contact over the entire structure (Figure S1).

\section{The Surgery}

The veterinary department's routine protocol for peri- and postoperative anaesthesia and analgesia management for neurosurgery was used. A median incision from the caudal part of the nasal bones to the external occipital protuberance was performed and the temporal muscles were dissected medially in order to expose the parietal bones. A patient tracker for neuronavigation was mounted on the skull over the nasion. A hole ($\diameter \SI{5}{\mm}$) was drilled into the parietal bone on both sides, each at the previously defined entry points in the direction of the planned track and the dura mater was incised. Both guide tubes were shortened to $\SI{27}{\mm}$ since a tip clearance distance of $\SI{9}{\mm}$ was needed for the adequate delivery of the four electrode contacts of the implantable Bioinduction Picostim--DyNeuMo-Mk1 $\SI{25}{\cm}$ DBS Leads (Bioinduction, Bristol, UK) to the CMN. Guide tubes were mounted from distal on the Renishaw\textsuperscript{\tiny\textregistered} neuroguide  instrument guide rod (Renishaw plc., Gloucestershire, UK), which had been fixed in the Scopis Biopsy Guide (Scopis GmbH, Berlin, Germany) and tracked for neuronavigation. A plastic slider was fixed on the rod immediately above the guide tube hub in such a way that the distal ends of both the tube and the rod formed a flush closure. The slider prevented displacement of the tube and a too deep advance of the rod into the brain during lowering of the assembly. Once the rod with the guide tube inside the biopsy guide was exactly aligned above the drill holes thanks to neuronavigation, the rod was lowered along the planned track into the brain parenchyma until the protruding edge of the guide tube hub met the surface of the bone and the rod was removed from the brain. Neuronavigation allowed for continuous supervision of the correct lowering of the guide rod. The section of the leads to be implanted was marked with a suture and inserted into the guide tubes. Leads and guide tube hub were fixed to the bone with bone cement and both proximal ends of the leads were inserted into the Bioinduction Picostim--DyNeuMo-Mk1 implanted DBS pulse generator (Bioinduction, Bristol, UK), which was attached to the frontal cranium with bone cement in a rubber sheath and fixed to the surrounding fascia with sutures. The muscle layer was sutured together over the stimulator and a routine closure was applied. Immediately after surgery CT imaging of the skull was performed in order to verify and compare the actual position of the guide tubes and leads with the pre-planned tracks.

\section{Statistical Analysis}

Where appropriate, the Mann Whitney U test was used in to compare seizure properties between the time periods before and after a given stimulation approach. Statistical analyses were performed with GraphPad Prism 9 (La Jolla, CA, USA). One-tailed tests were used and a $\mathrm{p} \le 0.05$ was considered significant.

\section*{References}
\bibliography{supplm}

\vspace{3.0\baselineskip}

\renewcommand{\thefigure}{S\arabic{figure}}
\setcounter{figure}{0}

\begin{figure*}[!h]
\centering
\includegraphics[width=115mm]{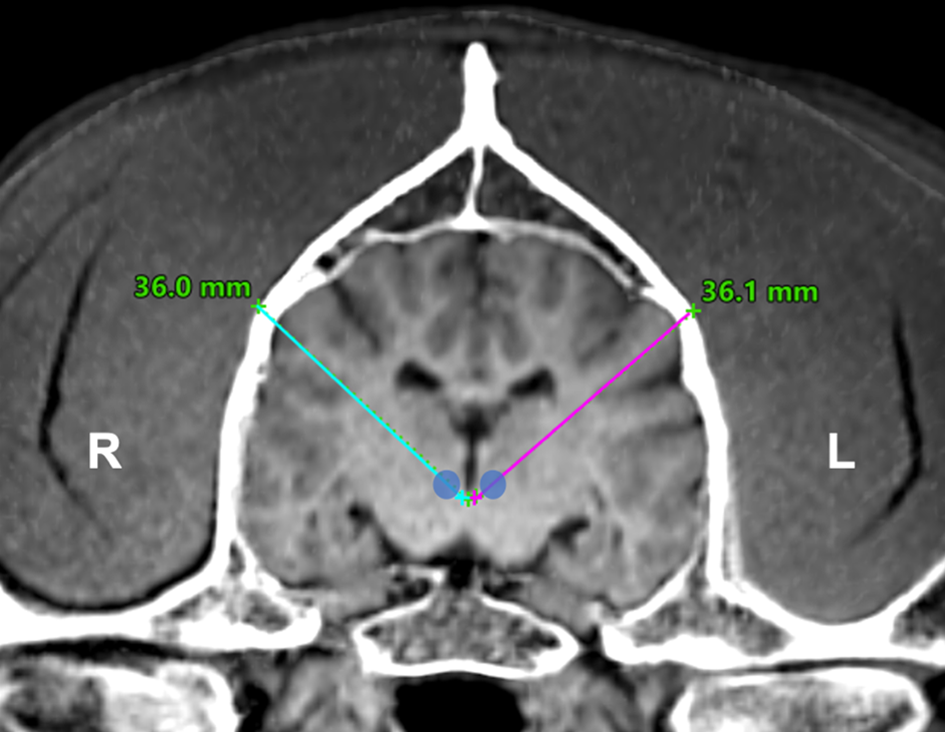}
\caption{Presurgical planning with transversal merged magnetic resonance imaging and computed tomography scans of the skull and brain with the Stryker CranialMap\textsuperscript{\tiny\textregistered} software. The blue (right hemisphere) and magenta (left hemisphere) trajectories represent the planned implantation sites and orientations of both guide tubes with the entrance points at the skull surface and the end points slightly ventromedially from the centromedian nuclei (CMN). The lengths of the planned trajectories are shown in green. Since a tip clearance of $\SI{9}{\mm}$ is needed for appropriate delivery of electrical information from the electrodes, guide tubes were shortened to $\SI{27}{\mm}$ each. Blue circles represent the CMN in both hemispheres.}
\end{figure*}